\documentstyle[epsfig,amsmath,amssymb,graphicx,soul,color,deluxetable,subfigure,lscape]{jaa}
\def\egs{\,erg~s$^{-1}$}      % note leading thinspace
\def\farcs{\hbox{$.\!\!^{\prime\prime}$}}

\def\gsim{\;\rlap{\lower 2.5pt\hbox{$\sim$}}\raise 1.5pt\hbox{$>$}\;}
\def\lsim{\;\rlap{\lower 2.5pt\hbo time lag between the formation x{$\sim$}}\raise 1.5pt\hbox{$<$}\;}
\def\la{\mathrel{\hbox{\rlap{\hbox{\lower4pt\hbox{$\sim$}}}\hbox{$<$}}}}
\def\ga{\mathrel{\hbox{\rlap{\hbox{\lower4pt\hbox{$\sim$}}}\hbox{$>$}}}}

\def\arcsec{\hbox{$^{\prime\prime}$}}

\def\farcs{\hbox{$.\!\!^{\prime\prime}$}}

\def\aj{AJ}% % Astronomical Journal
\def\apj{ApJ}% % Astrophysical Journal
\def\aa{A\&A}% % Astronomy and Astrophysics
\def\mnras{MNRAS}% % Monthly Notices of the RAS
\begin{document}
\title[X-ray Flares from young stars in the open clusters NGC 869 and IC 2602]{X-ray flares observed from six young stars located in the region of star clusters NGC 869 and IC 2602}

\author[Himali Bhatt et al.]{Himali  Bhatt$^{1}$,
J. C. Pandey$^{2}$,
K. P. Singh$^{3}$,
Ram Sagar$^{2}$, 
Brijesh Kumar$^{2}$ \\
$^{1}$Astrophysical Sciences Division, Bhabha Atomic Research Center, Trombay, Mumbai 400 085, India \\
$^{2}$Aryabhatta Research Institute of Observational Sciences, Manora Peak, Nainital 263 129, India\\
$^{3}$Tata Institute of Fundamental Research, Mumbai 400 005, India\\
}
%\ead{mshimali@gmail.com}
\pubyear{xxxx}
\volume{xx}
\date{Received xxx; accepted xxx}
\maketitle
\label{firstpage}

\begin{abstract}

We present an analysis of seven intense X-ray flares observed from six stars (LAV 796, LAV 1174, SHM2002 3734, 2MASS 02191082+5707324, V553 Car, V557 Car) for the first time. These stars are located in the region of young open star clusters NGC 869 and IC 2602. These flares detected in the XMM-Newton data show a rapid rise (10-40 minutes) and a slow decay (20-90 minutes). The X-ray luminosities during the flares in the energy band 0.3-7.5 keV are in the range of $10^{29.9}$ to $10^{31.7}$ erg s$^{-1}$. The strongest flare was observed with the ratio $\sim 13$ for count rates at peak of the flare to the quiescent intensity. The maximum temperature during the flares has been found to be $\sim$100 MK. The semi loop lengths for the flaring loops are estimated to be of the order of $\rm{10^{10}}$ cm. The physical parameters of the flaring structure, the peak density, pressure, and minimum magnetic field required to confine the plasma have been derived and found to be consistent with flares from  pre-main sequence stars in the Orion and the Taurus-Auriga-Perseus region.

\end{abstract}


\section{Introduction} \label{sec:st_int}
Stellar flares are events in which a large amount of energy is released in
a short interval of time. Flares are frequently observed from low mass ($<$2 M$_\odot$) active
stars throughout  the electromagnetic spectrum from radio to X-rays, e.g.,
H$\alpha$ (6563 \AA), X-rays (1.2-124 \AA) and radio (6-20 cm).
In a 13 days long observation of Orion (Chandra Orion Ultradeep Project), 
Favata et al. (2005) reported an average of 1.5 flares per star from young 
solar analogs. In a standard model, stellar flares are known to be a 
manifestation of the reconnection of magnetic loops, accompanied by particle 
beams, chromospheric evaporation, rapid bulk flows or mass ejection, and 
heating of plasma confined in loops (e.g., Sweet 1958, Parker 1957, Petschek 
1964, Yokoyama \& Shibata 1997, 1998, 2001, Priest \& Forbes 2002, Haisch 1989, 
G$\ddot{u}$del 2002). 
The flares produced by pre-main-sequence (PMS) stars are 
often detected  in the X-rays. These X-ray flares 
show extreme luminosities ($\geq10^{32}$ \egs) and very hot temperatures up to 
100 MK (e.g., Skinner et al. 1997; Tsuboi et al. 1998; Imanishi et al. 2002; 
Feigelson et al. 2002; Skinner et al. 2003; Preibisch et al. 2005). Wolk et al. 
(2005) studied the properties of flares of PMS stars in the Orion Nebula 
cluster and reported  the median peak luminosity of $\rm{10^{30.97}}$ \egs~  
with extremely hard spectra at peak time. An analogous study was presented by 
Stelzer et al. (2007) for PMS stars in the Taurus Molecular Cloud. The extreme 
flaring recorded on these PMS stars may have an important bearing on coronal 
heating. They are therefore providing observational evidence for the type of 
structures postulated by magnetospheric accretion models for PMS stars. Flares 
from PMS stars though present many analogies with the solar flares but they also
show significant differences, such as the amount of energy released. Therefore, analysis of the light curves during flares can provide insights into the 
characteristics of the coronal structures and ,therefore, of the magnetic field (Osten et al. 2010, Pandey \& Singh 2008, Reale et al. 2004).  

There is little observational evidence of flaring from massive ($>$10 M$_\odot$) and intermediate mass (2-10 M$_\odot$) stars. Only a few examples are known: 
these are $\lambda$ Eri (Smith et al. 1993), MWC 297 (Hamaguchi et al. 2000), 
$\sigma$ Ori E (Sanz-Forcada et al. 2004), and $\theta^2$ Ori A (Petit at el. 
2012). The presence of the high temperature plasma with kT $>$ 2 keV during the 
flaring activities in massive and intermediate mass stars cannot be explained 
in the standard framework for understanding X-ray emission from massive stars, 
i.e., wind shock model (Lucy \& White 1980, Owocki \& Cohen 1999, Kudritzki \& 
Puls 2000; Crowther 2007; Bhatt et al. 2010a,b), which predict the plasma temperatures of $<$ 1 keV.  
These flares are either explained by magnetically confined wind shock models (Babel \& Montmerle 1997) or
associated with nearby companion stars (Petit et al. 2012), but no convincing 
explanation has been forthcoming. 

In this paper, we report evidence of flaring activity in lightcurves of six 
stars which were detected in 
a timing analysis of the XMM-Newton observations of eight young open clusters 
described in our previous paper (Bhatt et al. 2013; hereafter Paper I).
The clusters were selected based on their ages ranging from 4 Myr to 46 Myr in 
order to bridge the gap between very young clusters like the Orion and 
relatively older clusters like the Pleiades to constrain the time evolution of 
X-ray emission. In Paper I, we have established the probable cluster membership 
of the X-ray 
sources on the basis of multi-wavelength properties of the X-ray sources and 
their location in color-magnitude diagrams of the cluster. Out of the eight 
young star clusters studied in Paper I, only NGC 869 and IC 2602 star clusters 
contain all 6 X-ray flare stars under present study. They are identified with 
the massive star BD+56 526 (B1III;Skiff 2009), intermediate mass stars  LAV 796 (B7;Currie et al. 2010 (ID\#1413)) and LAV 1174 (B9;Currie et al. 2010 (ID\#376)), and low mass stars  SHM2002 3734 (A7;Currie et al. 2010 (ID\#3212)), 
V553 Car and V557 Car (G0;Glebocki \& Gnacinski 2005), respectively.
The basic properties of these sources are given in  Table~\ref{tab:basic}, along with relevant information about their host clusters. Though the region of 
both star clusters NGC 869 and IC 2602 were observed earlier in X-rays by 
Currie et al. (2010) and by second ROSAT PSPC Catalog 2000 respectively, the
timing characteristics of these sources were not studied. In fact,  X-ray flare activity from these six sources are reported here for the first time.  
Their observed V magnitudes and (B-V) colors are given in Table~\ref{tab:basic}.
Their photometric spectral types have been derived on the basis of their intrinsic colors after correcting their extinction towards their corresponding clusters, i.e., $(B-V)_0$=(B-V)-E(B-V), and given in Table~\ref{tab:basic}. 
The spectroscopic information about these sources has been taken from Vizier\footnote{ http://vizier.u-strasbg.fr/viz-bin/VizieR} services.

The present study provides a valuable sample of X-ray flares from stars in 
different mass groups to investigate the possible mechanisms for generation of 
X-ray flares. In addition, the morphologies of these flares are also derived 
along with a general comparison of the characteristics of these flares 
with those of other X-ray active stars and the Sun. In the following section,   
detection of X-ray flares and their possible association with massive, intermediate and PMS low mass stars has been explored.  The characteristics of flares 
are defined in section 3. The physical parameters of coronal loop structures 
have been modeled in section 4 while results are discussed and concluded 
in section 5.

\section{Detection of X-ray flares}
\label{sec:cross}
\begin{figure*}
\includegraphics[width=2.45in, height=2.45in]{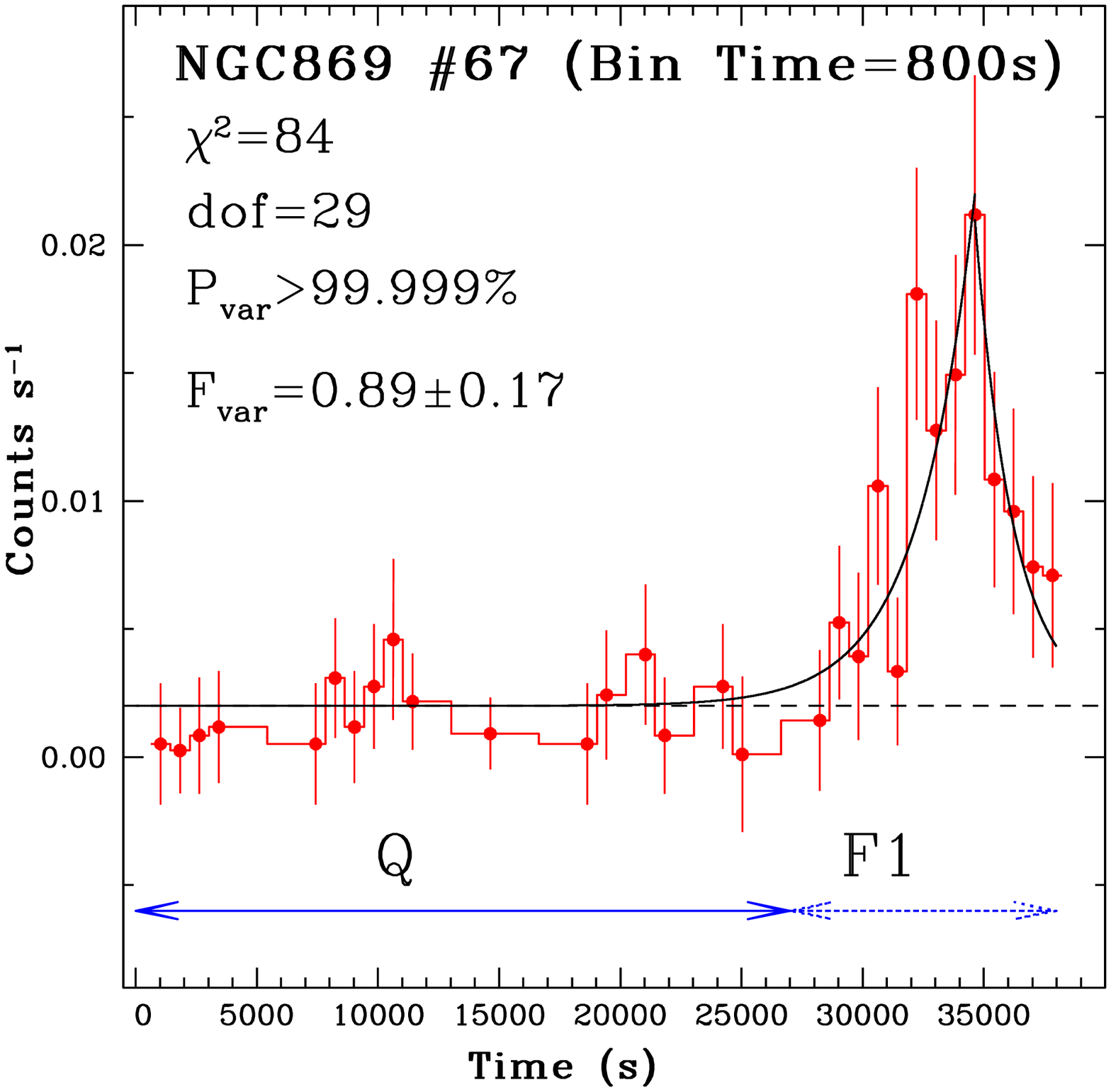}
\includegraphics[width=2.45in, height=2.45in]{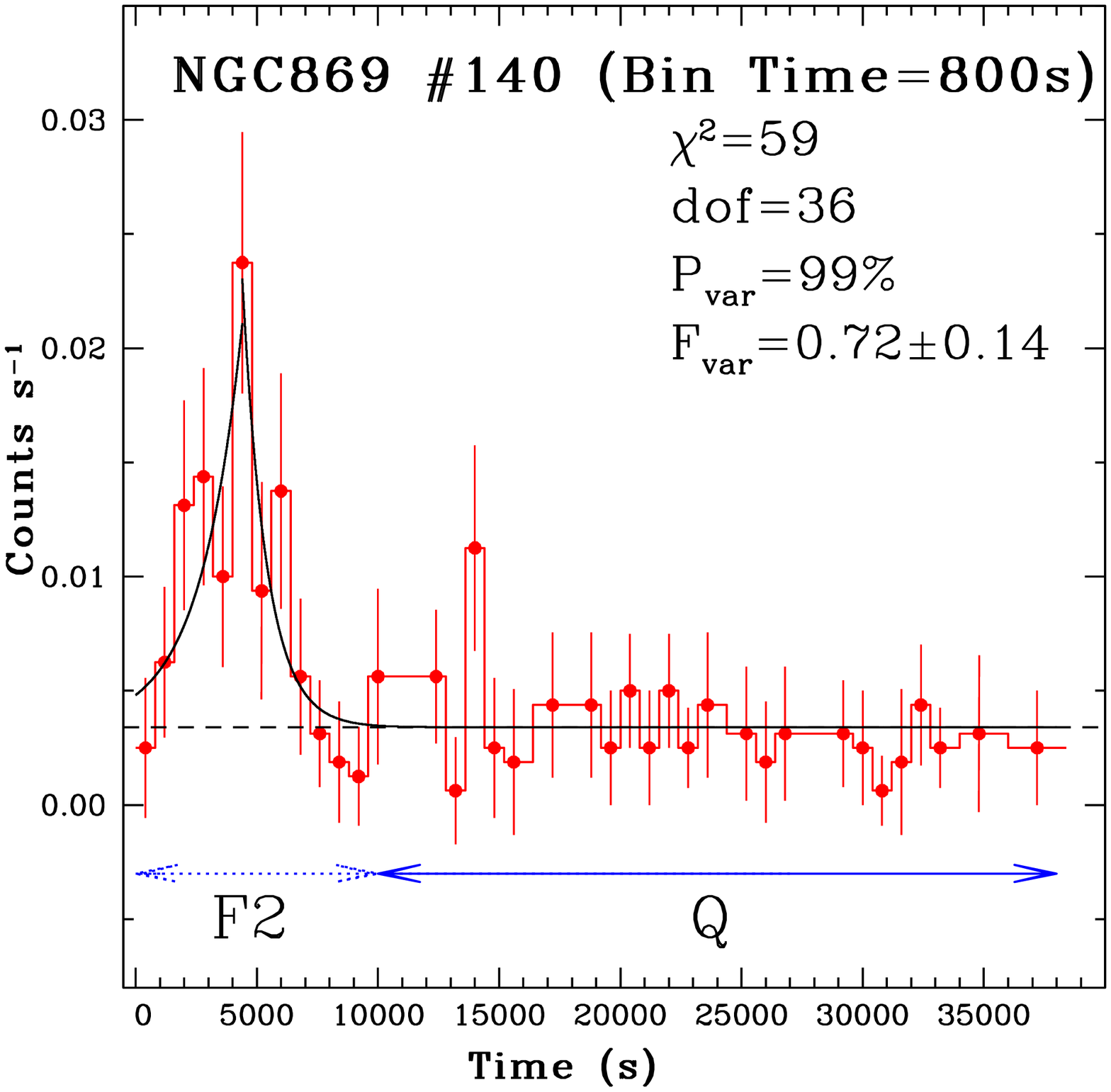}
\includegraphics[width=2.45in, height=2.45in]{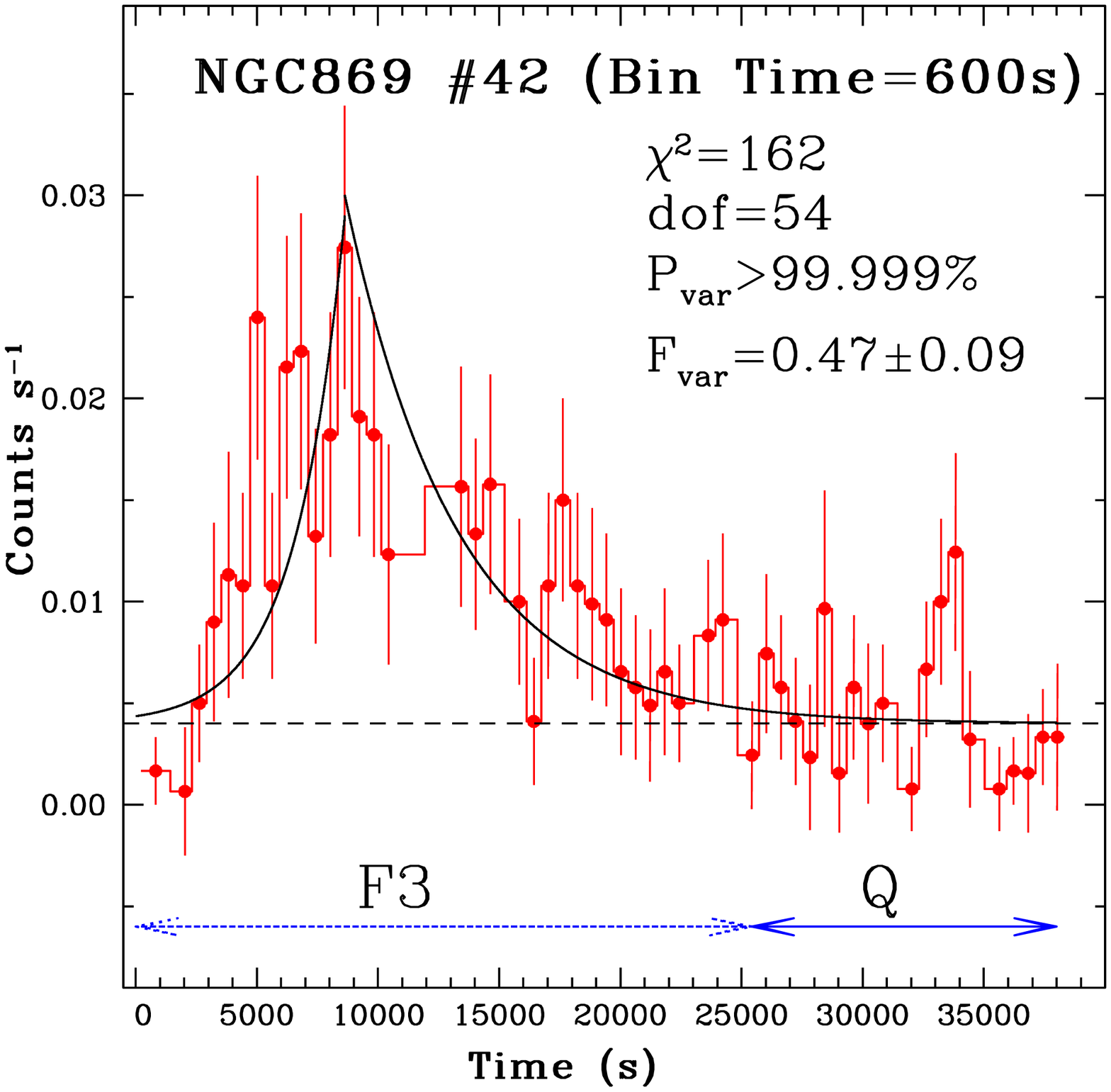}
\includegraphics[width=2.45in, height=2.45in]{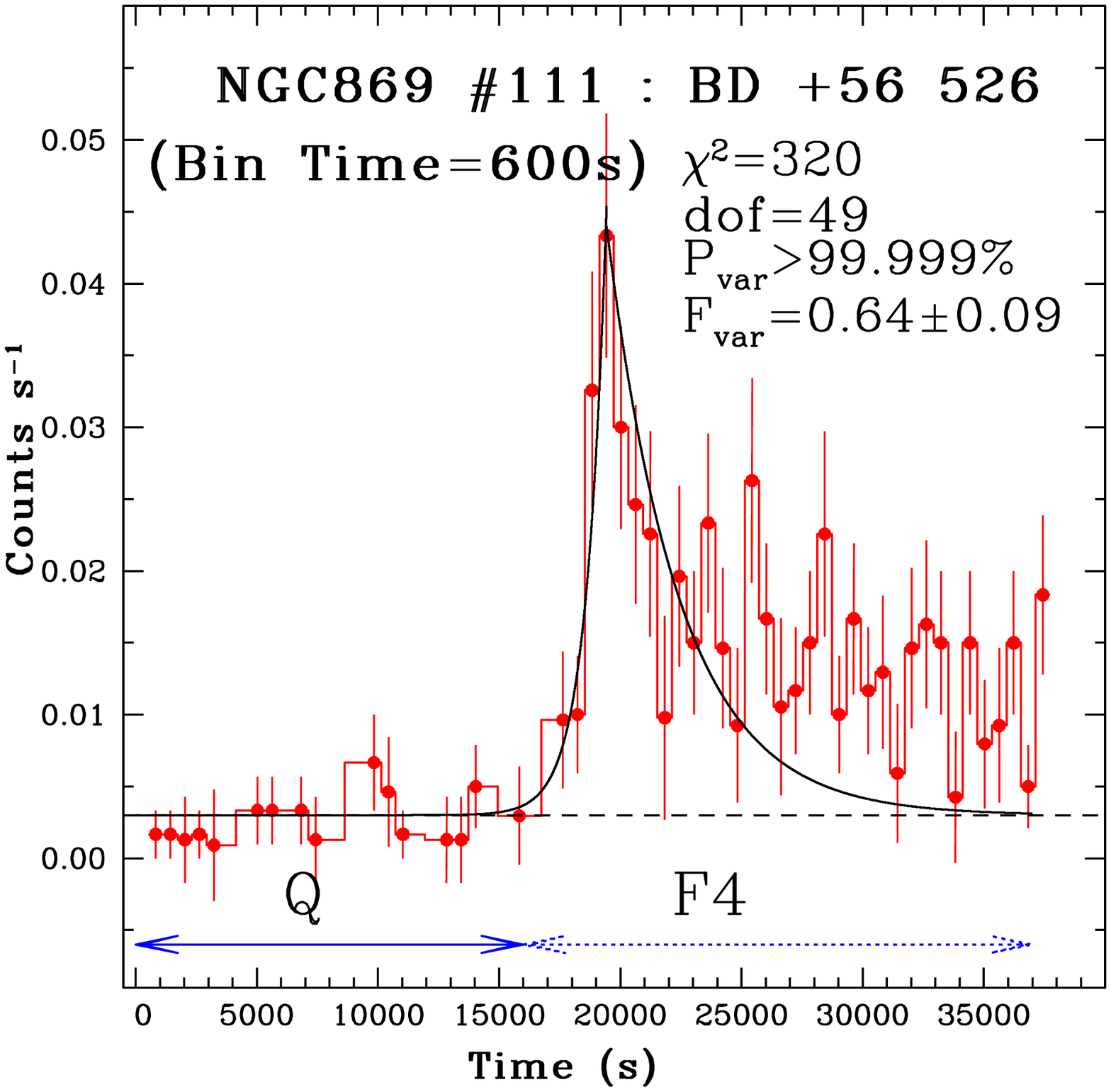}
\includegraphics[width=2.45in, height=2.45in]{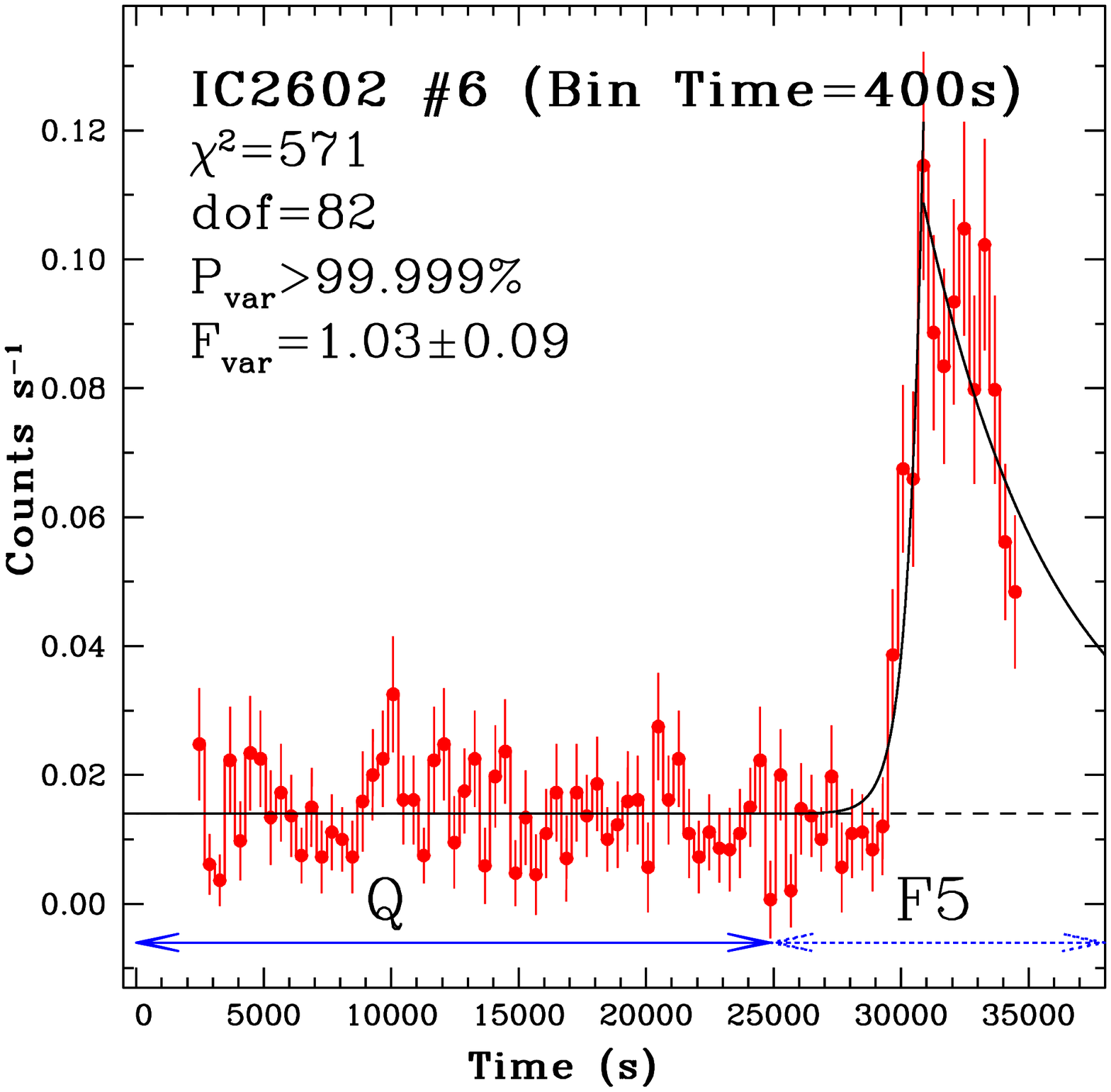}
\includegraphics[width=2.45in, height=2.45in]{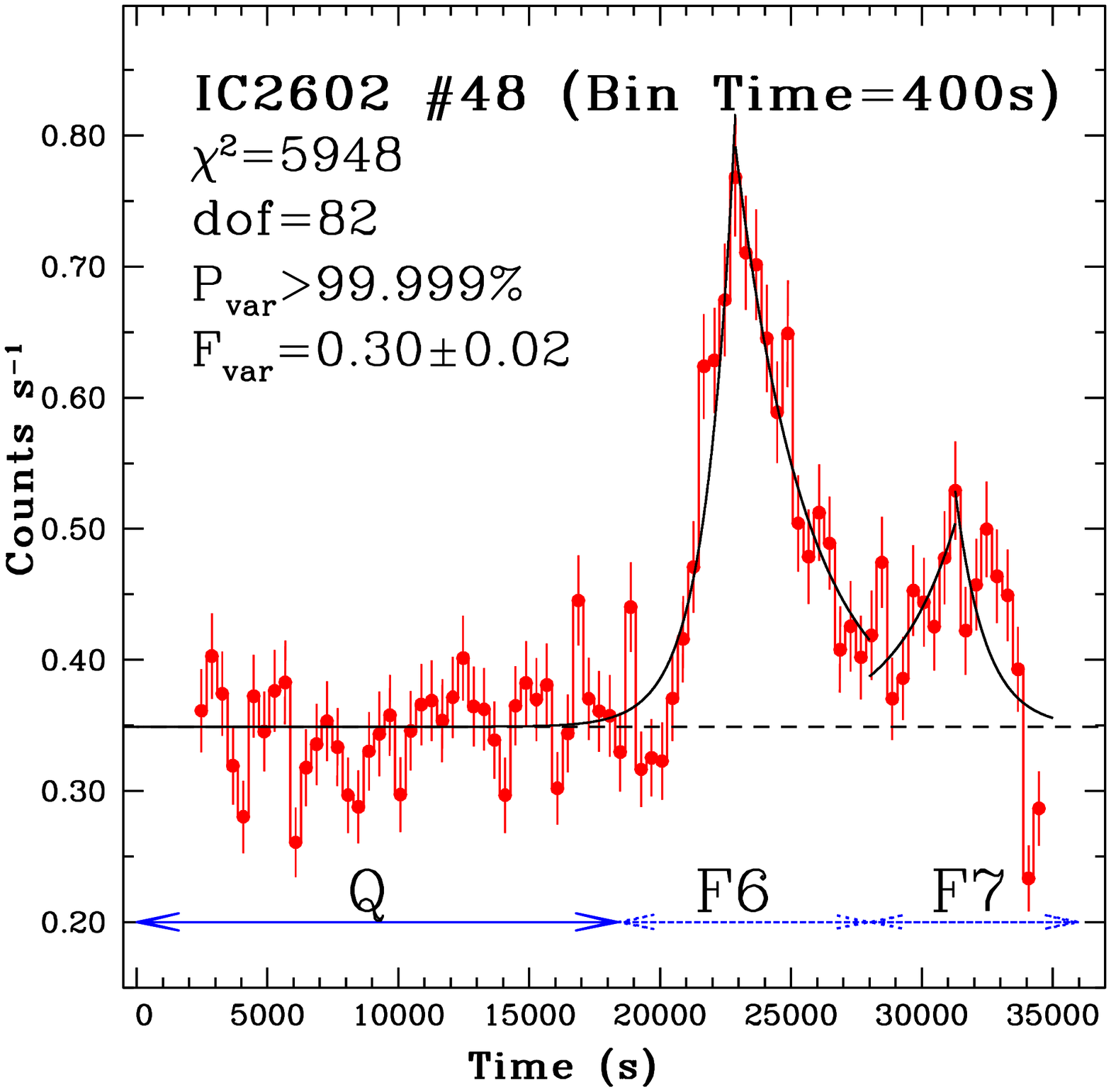}
  \caption{X-ray flares from the cluster members. Flare time intervals are represented by dotted arrows with lines and
marked by F$\it{i}$ , where $\it{i}$ =1,2,...,5 refers to the flare number. Quiescent state 'Q' time intervals
 are shown by solid arrows with lines and quiescent state mean count rates are marked with dashed lines. Exponential rise and decay 
fit are shown by solid lines.
 }
  \label{fig:flare}
\end{figure*}

The details of observations, data reduction and identification procedure of the X-ray sources within Two-micron all sky survey (2MASS) Point Source Catalog (PSC; Cutri et al.  2003) and Optical spectroscopic catalogues of stars from 
 Webda\footnote{http://www.univie.ac.at/webda/navigation.html}  and Vizier\footnote{ http://vizier.u-strasbg.fr/viz-bin/VizieR} are given in Paper I where each 
X-ray source detected in  data reduction procedure has been ascribed a unique  
identification number (ID). The IDs of X-ray sources (in Paper I) with flare 
like features are given in Table~\ref{tab:basic}.   The light curves and 
spectra of the stars within the open clusters have been extracted using 
circular extraction regions with radii varied between 8$^{\prime\prime}$ and 
40$^{\prime\prime}$ centered on the source position in the
energy range of 0.3--7.5 keV. The background data have been taken from several neighboring source-free regions on the detectors. Data have been binned from 400 
s to 1600 s. X-ray light curves of stars showing flaring features are shown in 
Figure~\ref{fig:flare}, where the ‘flare regions’ have been represented by arrows and marked by F$\it{i}$, where ${\it{i}}$ = 1, 2, . .  ,7 refers to the flare number (FN), and the quiescent regions have been marked as 'Q'. 
The probabilities of the variability in the X-ray lightcurves were derived using $\chi^2$ test and are given in Figure~\ref{fig:flare}.  
In the X-ray light curves, flares are characterized by two or more 
consecutive time bins that constitute a sequence of either rising or falling 
count rates, corresponding to rise and decay phase of the flare. A rise 
immediately followed by a decay with a peak count rates greater than 3$\sigma$ 
of the count rates in quiescent regions 'Q' is counted as one flare. 
In this way seven flares ,F1 to F7, were detected. Except from star IC 
2602 \# 48, only one flare event was detected from each star.  

\subsection{Massive and Intermediate mass stars}
\label{sec:demass}
Three flares F4, F1 and F2  are suspected to be related to massive star BD+56 526 and intermediate mass stars
LAV 796 and  LAV 1174, respectively. Therefore, we have carefully examined these X-ray sources to check for the possibility of X-ray emission from 
presumed nearby/companion star since most of the flare stars have an optical/NIR counterpart within 3\arcsec~ (see paper I), except  BD+56 526. Flare F4 from 
the X-ray source \# 111 of cluster NGC 869 is detected from a position that 
has an offset of 6\arcsec~from the  massive star BD+56 526.  

\begin{figure*}
\centering
\subfigure[]{\includegraphics[width=70mm]{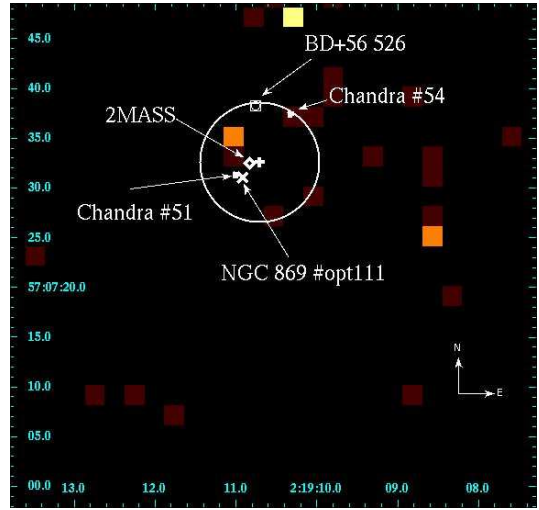}}
\subfigure[]{\includegraphics[width=70mm]{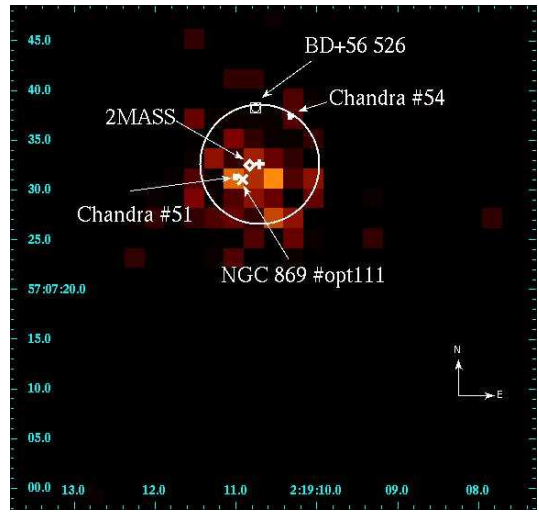}}
\caption{X-ray image of the field BD+56 526 (a) before flaring event and (b) during
the flaring event.
 X-axis and Y-axis are representing RA(J2000) and DEC(J2000), respectively.
The {\sc Xmm-Newton} detection position is marked by the symbol of plus.
The symbol of the square represents {\sc Chandra} source detection position  of
source ID \#51 and \#54 from Currie et al. (2009), symbol of diamond represents 2MASS H-band detection position and square with circle represent the position of massive star BD+56 526. The optical position given by Mayne et al. (2007) is marked by the symbol of the cross.}
\label{fig:BD_image}
\end{figure*}

Because of the large offset between the position of BD+56 526 and that of  X-ray source \# 111 in NGC 869, 
the association of X-ray source with the massive star BD+56 526 is doubtful. 
This X-ray emission  could be either from the B-type massive star, BD+56 526, or a presumed T-Tauri star located close to BD+56 526
 but unresolved in the present observations.
In order to check whether the position had shifted between the flaring episode and the quiescent interval, we have extracted   
X-ray images of the source  before and during the flaring activity. These images  are shown in
Figure \ref{fig:BD_image}.
During the quiescent state no X-ray emission has been
detected within a circle of 6\arcsec~radius around the X-ray source position during the flare. This implies that
the X-ray emission is  dominated by the flare.
Further, the optical and 2MASS NIR counterparts of the  NGC 869\#111 have been searched within a radius of
6\arcsec~about the X-ray source in the available catalogues. No optical detection
other than BD+56 526 has been seen with in USNO B-1.0 catalogue down to $\rm{V=21}$ mag.
However, an optical counterpart (V=19.401 mag) has been detected with an offset of 2\farcs3 in the recent
photometry carried out by Mayne et al. (2007). For convenience, we name this source as NGC 869\#opt111.
This optical position is marked by a cross  symbol in Figure~\ref{fig:BD_image}.
In the 2MASS catalogue, one source, 2MASS 02191082+5707324,
has been detected only in  the H-band (H$=$15.274$\pm$0.454 mag)
 which is marked by a diamond in Figure \ref{fig:BD_image} with an offset of 1\arcsec~from  NGC 869\#111.
 The offset between  2MASS 02191082+5707324
 and the optical counterpart is 1\farcs6.
Recently, X-ray data from {\sc Chandra}
mission have been published by Currie et al. (2009) for the cluster NGC 869.
We have searched for the corresponding  counterpart of   NGC 869\#111 in the {\sc Chandra} X-ray source catalogue,
and found two sources within  6\arcsec~search radius.
These sources are identified with
{\sc Chandra} NGC 869\#51 and {\sc Chandra} NGC 869\#54 (Currie et al. 2009)
and marked in Figure~\ref{fig:BD_image} using squares. They show an offset of 2\farcs6 and 5\farcs7, respectively, from the
X-ray source  NGC 869\#111.
The offset of the optical and the NIR detection from the {\sc Chandra} NGC 869\#51 are 0\farcs4 and 1\farcs8, respectively.
All the sources, namely,  NGC 869\#opt111,  2MASS 02191082+5707324 and {\sc Chandra} NGC 869\#51 are located
towards the south-west direction of   NGC 869\#111 (Figure~\ref{fig:BD_image}).
It appears that the XMM-Newton detection is most likely associated with one of these sources.
It is also quite possible that all the four objects identified with optical ( NGC 869\#opt111), 2MASS ( 2MASS 02191082+5707324), 
Chandra (NGC 869\#51) and XMM-Newton (NGC 869\#111) catalogues, are from a single source as their offsets are very small.
 It also suggests  that the XMM-Newton detection ( NGC 869\#111) is not likely to be associated with the star BD +56 526.
Furthermore, the X-ray luminosity of BD+56 526 has been predicted  to be of the order of $\rm{10^{31}}$ \egs ~using the relation
from Bergh$\rm{\ddot{o}}$fer et al. (1997) and Sana et al. (2006), whereas the X-ray luminosity during quiescent state is  found  to be  $\rm\sim{10^{30}}$ \egs,
based on the mean count rate of 0.003 $\rm{counts~s^{-1}}$ during the quiescent state and the corresponding flux estimated using
WebPIMMS\footnote{http://heasarc.gsfc.nasa.gov/Tools/w3pimms.html} with {\sc apec} model
for 1 keV plasma.
Therefore, we do not believe we have detected X-rays from BD+56 526 directly, and 
the flare-like feature from the neighbourhood of BD+56 526 is probably associated with a nearby low mass star  
which is not resolved due to the poor resolution of XMM-Newton.

The photometric spectral type of the source  NGC 869\#111 (2MASS 02191082+5707324) has been estimated
assuming that it is a  member of cluster NGC 869 with optical and NIR counterparts.
 Using the distance (2.3 kpc) of the cluster NGC 869
and the value of V = 19.401 mag from Mayne et al. (2007) photometry, the absolute magnitude M$_V$ of XMM-Newton source is
estimated to be nearly 5.86 mag. The observed (V-H) colour of the  XMM-Newton source is found to be nearly 4.13 mag.
Using the reddening towards the cluster E(V-H)=1.467 mag,
the intrinsic color of the XMM-Newton source is derived to be 2.66 mag.
The intrinsic (V-H) color and absolute V band magnitude are consistent with 
those for a K-type low mass star (Landolt-B$\ddot{o}$rnstein 1982; Ducati et al. 2001),
and at the age of 13 Myr, for the cluster NGC 869,  this X-ray source could be a low mass PMS star. 
The uncertainty in the V-band is not given in the catalogue by Mayne et al. (2007), therefore the 
errors in absolute magnitude and colors cannot be determined.
However, the large error in the H-band magnitude produces an error of two to three subclasses in the determination of the spectral type.

Currie et al. (2009) have reported high spatial resolution observations of the open cluster NGC 869
with the {\sc Chandra} X-ray observatory. In their catalogue, X-ray source with identification number
\#182 is found to be only 3\arcsec~ away from our X-ray position of LAV 1174.
The offset between its 2MASS NIR position and {\sc Chandra} positions is  found to be 0.3\arcsec~.
No other X-ray sources were detected in {\sc Chandra} data within 6\arcsec search radius of LAV 1174, spatial resolution limit of XMM-Newton. 
Therefore, it is quite likely that the X-ray flare F2 detected by us is associated with X-ray source \#182 with {\sc Chandra} observations.
The X-ray source \#182 in {\sc Chandra} catalogue was associated with an intermediate mass star \#376 (LAV 1174) 
in an optical spectroscopic catalogue reported by Currie et al. (2010). 
No X-ray source has been detected within  6\arcsec search radius of our X-ray position of LAV 796 in the open cluster NGC 869 in  {\sc Chandra} data.
Thus, X-ray flares F1 and F2 appear to be related to intermediate mass stars LAV 796 and LAV 1174, and
we are not able to rule out the possibility that X-ray flares arise directly from intermediate mass stars.

\subsection{Low mass stars}

Three X-ray flares F3, F5 and F6 are associated with low mass stars  SHM2002 3734,  V553 Car and V557 Car, respectively.
X-ray emission from SHM2002 3734 was observed by {\sc Chandra} from a source with 
an identification number  \# 255 in Currie et al. (2009) catalogue and
has an offset of only 3\arcsec~ with our X-ray positions.  
Two X-ray sources J104100.6-642003 and J104241.3-642106 with an offset of about 3\arcsec~  from V553 Car 
and V557 Car, respectively have been reported in the second ROSAT PSPC Catalog (2000). 
These two sources V553 Car and V557 Car  in the open cluster IC 2602 are probably responsible for the three flares
F5, F6 and F7  and characterised as BY Dra-type variables in the 76th name list of variable stars (Kazarovets, Samus \& Durlevich 2001)"

\section{X-ray flare characteristics}
\label{sec:flare_char}

\subsection{Rise time and Decay time}
\label{sec:rsdec}
The lightcurves of the flares (see Figure~\ref{fig:flare}) are characterised by a 
fast rising phase followed by a slower decay phase, which are similar to the solar flares and thus solar-like magnetic reconnection events which govern
by conduction and radiation cooling mechanisms. This fast-rise exponential 
decay shape (usually called FRED) are commonly associated with stellar flares 
from PMS stars and fitted with exponential function (see Stelzer, 
Neuh$\ddot{a}$user \& Hambaryan 2000, G$\ddot{u}$del 2004 ). The e-folding rise and decay times have been determined from the least-squares fit 
of the exponential function 

\begin{equation}
\rm { c(t) = A_0 exp^{-[(t-t_0)/\tau_{r,d}]} + q   }
\label{expfun}
\end{equation}

where $c(t)$ is the count rate as a function of time $\it t$, $\it {t_0}$
is the time of peak count rate, $\it q$ is the count rate in the quiescent
state, $\it \tau_d/\tau_r$ is the decay/rise time of the flare and $\it A_0$ is the count
rate at flare peak. The quiescent state count rates were taken as mean value of count rates during the 'Q' in the light curves. 
The least-square fits of the eq. \ref{expfun} are  shown in Figure~\ref{fig:flare} by solid lines.
The values of $\tau_d$ and $\tau_r$ along with the start time, flare duration, quiescent state count rates and ratio of count rate at flare and quiescent state are  given in Table~\ref{tab:flare_par} for all the seven flares observed.
The rise time of these flares were found in the range of 10-40 minutes, whereas the e-folding decay times were found in the range of 20-90 minutes.
The  longest duration ($\sim$ 25 ks) flare was observed from the star \# 42 of 
NGC 869 (SHM2002 3734). The peak flare count rates were found to be 1.5-13.3 
times more than that found during quiescent state. The strongest flare was 
observed from the star \# 111 of NGC 869 (2MASS 02191082+5707324), where the 
ratio of peak to the quiescent count rate was found to be 13.3.

\subsection{X-ray spectra of flares}
\label{sec:spec}
X-ray spectra of the flares have been analysed to look for the spectral evolution during the flares.
Spectral analysis has been performed based on global fitting using
the Astrophysical Plasma Emission Code ({\sc{apec}}) version 1.10 modeled by Smith et al. (2001) and
 implemented in the {\sc xspec} version 12.3.0.
The absorption toward the stars by interstellar medium was accounted for by using
multiplicative model photoelectric absorption screens ({\sc phabs}) in {\sc xspec} (Baluci$\acute{n}$ska-Church \& McCammon 1992).
In order to study the flaring component only, we have performed "{\sc{2T apec}}" model spectral fits of the data during the flare state.
The quiescent emission taken into account 
by including its best-fitting 1T model as a frozen background contribution. 
This is equivalent to considering the flare emission after subtraction of the quiescent level, 
and allows us to derive the effective average temperature and the emission measure (EM) of the flaring plasma. 
The spectral parameters and coronal temperatures for quiescent state (Q) are derived in paper I, 
and the values of hydrogen column density ($N_H$) and abundance ($Z$) parameters have been kept fixed to the quiescent state values in the spectral fitting (see Table~\ref{tab:flare_fit_par}). 
The best-fit parameters during the flare are given in Table~\ref{tab:flare_fit_par}. The X-ray luminosity during the flares varied from $10^{29.9}$ to $10^{31.7}$ erg s$^{-1}$.  For the flare F4 from LAV 796, the X-ray luminosity during the flare was found to be   $10^{31.7}$, which is 16.6 times more than that found during the quiescent state.
According to Reale (2007), a typical output of the analysis of X-ray spectra with moderate resolution, 
such as those from CCD detectors (e.g. EPIC/XMM-Newton), is the average 
temperature of the flaring  loops which is usually lower than the real loop maximum temperatures. 
The loop maximum temperature for the EPIC instruments are calibrated as $\rm T_{max} = 0.13~T^{1.16}_{obs}$ (Reale 2007).

\section{Modeling the X-ray flares based on coronal loops}
\label{sec:loop_model}
The characteristics of these flares are all consistent with the flares from low mass PMS stars 
in young open clusters, e.g., the Orion (Getman et al. 2008a), and
therefore, can be associated with the presence of a magnetically confined corona and thus of an operational dynamo mechanism. 
The physical sizes and morphology of the loop structures involved in a stellar flare can 
be estimated using flare loop models by an analogy with solar flares.
However, it has been considered that flares occur 
inside closed coronal loops anchored to the photosphere, plasma 
is confined inside each loop, and  its bulk motion and 
energy transport occur only along the magnetic field lines ( Reale et al. 1997).
 
A method to infer the geometrical sizes and other relevant physical
parameters of the flaring loops was presented by Reale et al. (1997),
 which was based on the decay time, and on evolution of temperature and the EM during the flare decay.
 However, this method needs time resolved spectroscopy, therefore, could not be used for the present analysis. 
Haisch (1983) assumed  that $\rm{\tau_d}$ is comparable to the radiative
and conductive cooling times, and estimated the loop length ($\rm{L_{Ha}}$) as,

\begin{equation}
\rm L_{Ha} = 5\times10^{-6} EM^{0.25} \tau_d^{0.75}
\label{len_Ha_cal}
\end{equation}
 Hawley et al. (1995) described an approach, which is  based on rise and decay times.
In this approach, strong evaporative heating is
dominant during the rising phase, while X-ray emission is dominated by
radiative cooling during the decay phase. At the loop top, there is an equilibrium  and the loop length can be derived as

\begin{equation}
\rm{
L_{r} = \frac{1500}{  (1-x_d^{1.58})^{4/7}  }.\tau_d^{4/7}.\tau_r^{3/7}.(T_{max})^{1/2}
}
\label{loop_ri_de}
\end{equation}

\noindent
where $\rm {T_{max}}$ is temperature
at flare apex and $\rm{ x_d^2 = q/A_0}$.
We have sufficient information to model a loop by rise and decay method and Haisch approach. 
The loop-lengths of the flares were derived in the rage of $1.3 - 3.8 \times 10^{10}$ cm using 
the rise and decay method, while it was $1.8-9.8 \times 10^{10}$ cm using Haisch approach. 
In general, the loop lengths derived from Haisch approach are found to be more than that derived from rise and decay method. 
However, in most of the cases the differences between loop lengths derived from these two 
methods are well with $1\sigma$ level. The flare F3 from the star SHM2002 3734 in the open cluster NGC 869 is 
only the flare where the difference between loop lengths derived from two methods  is more than $2\sigma$ level.

After knowing loop lengths, the various loop parameters like maximum pressure, plasma density, 
loop volume and minimum magnetic field to confine the plasma  can be derived.
The analysis of X-ray spectra provides values of the temperature and EM.
From the EM and the plasma density ($n_e$), the
volume (V) of the flaring loop is estimated as $EM = n_e^2$ V.
Using the loop scaling laws, the maximum
pressure, temperature, loop length  and heating rate per unit volume are related as (Rosner, Tucker \& Vaiana 1978, Kuin \& Martens 1982):

\begin{equation}
T_{max}  = 1.4 \times 10^3 (pL)^{1/3}
\label{eq:pressure}
\end{equation}

\begin{equation}
E_H \approx 10^{-6} T_{max}^{3.5} L^{-2}  ~~~ergs ~s^{-1} cm^{-3}
\label{eq:EH}
\end{equation}

\noindent
The minimum magnetic field necessary to confine the flaring plasma can be
estimated as
$B = \sqrt{8 \pi p}$.

The derived parameters are given in Table \ref{tab:loop_length}. 
Using the equation \ref{eq:pressure},
the maximum pressure was found to be in the range of $5\times 10^2 - 8.75\times 10^3$ dyne cm$^{-2}$ during the flare. 
Assuming that the hydrogen plasma is totally ionized, the maximum plasma density in loop during the flare was found from $5\times 10^{10}$ to $1.5\times 10^{12}$ $\rm{cm^{-3}}$.
 Using the observed EM, the loop volume was computed in the range of $1.5\times10^{30} - 5.7\times10^{32}$ cm$^3$. 
A hint for the heat pulse intensity comes from the flare maximum temperature. By applying
the loop scaling laws (see eq. \ref{eq:EH}) and loop maximum temperature (see Table \ref{tab:loop_length}), the heating rates per unit volume ($E_h$) are
 found from 0.3 to 92 erg s$^{-1}$ cm$^{-3}$. From the derived pressure of the flare plasma, minimum magnetic field required to confine the plasma are found to be in the range of 0.1-1.5 kilo Gauss.   

\section{Discussion and Conclusions}
\label{sec:con}

We have carried out analysis of seven flares observed from six stars located in the region of young star clusters NGC 869 and IC 2602. Out of six stars, four 
have been classified as low mass stars, while two stars LAV 796 and LAV 1174 
have been classified as intermediate mass stars. The strongest flare was 
observed from the star \# 111  of NGC 869 with a ratio of peak flare to 
quiescent state count rates of $\sim 13$.  It is detected from the 
neighbourhood of BD+56 526 
and is probably associated with a nearby low mass star, 2MASS 02191082+5707324, which is not resolved due to the poor resolution of XMM-Newton. 

The rise and decay times  of the flares reported here are lower than that of the flares from PMS stars in Orion (between 1 hour to 1 day;Getman et al. 2008a), 
but, compared to that of solar flares (e.g., only a few minutes; Priest \& Forbes 2002), 
PMS stars of Pleiades cluster (Stelzer, Neuh$\rm{\ddot{a}}$user \& Hambaryan 2000) and 
main-sequence (MS) stars (Pandey \& Singh 2008).  
The peak to quiescent state count rate ratio for these flares are in the range of 1.5 to 13.3.
The maximum flare plasma temperatures are found to be  comparable to the flares from PMS, MS  and RCVn-type stars 
(e.g., Pandey \& Singh 2008, 2012; Getman et al. 2008b; Favata et al. 2005).  
The total energy released during the X-ray flares (= $\rm{L_X \times duration ~of ~flare}$), has been found to be in the range of 10$^{34.0}$ - 10$^{35.8}$ ergs. 
This indicates that these flares are 10$^{4-6}$ times more energetic than the solar flares (Moore et al. 1980), but similar to the flares observed 
from field dwarfs, and evolved  RS CVn-type binaries (e.g., Pandey \& Singh 2008, 2012).
The inferred sizes of the flaring structures are consistent with the results presented by
Getman et al. (2008b) and Favata et al. (2005)  for certain PMS stars.
However, the derived  values of the maximum plasma temperatures and loop lengths are
larger than that of the solar flares. The giant X-ray emitting arches found in the Sun are to have typical sizes of
$\rm{10^8}$ cm with the maximum temperature of 10 MK
(e.g., Getman \& Livshitz 1999;Getman \& Livshits 2000).
These parameters are similar to the parameters derived for the PMS stars in Taurus-Auriga-Perseus region (Stelzer, Neuh$\rm{\ddot{a}}$user \& Hambaryan 2000)
and the Orion (Getman et al. 2008b; Favata et al. 2005).

X-ray luminosities ($\rm{L_X}$) during the  flares from low mass stars SHM2002 3734 (\# 42) in the cluster NGC 869, and
V553 Car  and V557 Car  in the cluster IC 2602  have been found to be of the order of $\rm{10^{30-31}}$ \egs.
It indicates that these flares are more powerful than those observed from
field dwarf stars (Pandey \& Singh 2008) but are equally powerful as the flares from PMS stars (Getman et al. 2008b). 
The loop length of the order of 10$^{10}$ cm is indicating large coronal structures. 
For these flares, the estimated maximum electron densities under
the assumption of a totally ionized hydrogen plasma  have been found in the
order of 10$^{10 - 12}$ cm$^{-3}$. It is comparable to the values expected
from a plasma in coronal conditions (Landini et al. 1986).
The strength of the magnetic field (of the order of a few hundred G), loop sizes of the large-scale magnetic field
and  X-ray luminosities of the flares suggest that these flares are enhanced analogs of eruptive solar flares.
Therefore, these flares can be interpreted as scaled-up versions of solar flares and
the flares from PMS stars in the Orion and the Taurus-Auriga-Perseus region. 

The loop lengths of the flaring structure from  
 intermediate mass stars LAV 796  and LAV 1174  in the cluster NGC 869 
have been found to be comparable with that of flares from 
low mass stars in the present sample and also in other studies  (e.g., Getman et al. 2008b; Favata et al. 2005),
but larger than that of the solar flares 
(e.g., Getman \& Livshitz 1999; Getman \& Livshits 2000). 
However, these values of loop lengths are lower that of the flares from young  massive stars  MWC 297 ( $\rm{10^{11}}$ cm; Hamaguchi et al. 2000)
and HD 261902 ($\rm{10^{12}}$ cm;Yanagida et al. 2005).
Similarities between the nature of X-ray flares from intermediate mass stars and low mass stars 
suggest that X-rays from  intermediate mass stars are probably arising from 
unresolved nearby low mass stars (Stelzer et al. 2006, Joshi et al. 2008 and references therein). 
Recently, Balona (2013) has studied 875 A-type stars using Kepler data in optical band, and
reported that variation in light could be due to rotation modulation caused by star-spots.
 If A-type stars have spots, then it is natural to expect a magnetic field, and therefore the possibility of X-ray activity and magnetic reconnection 
in intermediate mass stars.
If the X-ray emission indeed arises from the intermediate mass stars 
rather than a hidden low mass star through coronal loop structures then the origin of the 
magnetic field of the order of 1000 Gauss is hard to explain in the intermediate mass stars which are thought to be 
fully or nearly-fully radiative.

% sec:acknowledgments
%_____________________________________________________________________

\section*{Acknowledgments}

Authors are thankful to anonymous Referee for providing constructive 
comments which improved content of the manuscript.
This publication makes use of data from the Two Micron All Sky Survey, which is a joint project of
the University of Massachusetts and the Infrared Processing and Analysis Center/California
Institute of Technology, funded by the National Aeronautics and Space Administration and
the National Science Foundation, and data products from XMM-Newton archives using the high 
energy astrophysics science archive research center 
which is established at Goddard by NASA. 
We acknowledge XMM-Newton Help Desk for their remarkable support in X-ray data analysis.  
Data from Simbad, VizieR catalogue access tool, CDS, Strasbourg, France have also been used. 
HB is thankful for the financial support for this work through the INSPIRE faculty fellowship granted by the Department of Science \& Technology  India, and
 R. C. Rannot,  Nilesh Chouhan and  R. Koul for their support to finish this project.

\clearpage

%%%%%%%%%%%%%%%
% Change back to the regular baselineskip, if necessary
%%% FIGURE %%%

\clearpage

%*****************

%*****************

\clearpage
%%%%%%%%%%%%%%%
% Tables
%%%%%%%%%%%%%%%

% tab:Basic properties
  %_________________________________________________________________
  \begin{table*}
\caption{Basic properties of stars with flare-like features.}             % title of Table\label{table:1}      % is used to refer this table in the text
\label{tab:basic}      % is used to refer this table in the text
\scriptsize
\begin{tabular}{ccccccccccc}       % centered columns (4 columns)\hline                 % inserts double horizontal lines
\hline 
Cluster  \#ID$^\dagger$  & Name         & $RA_{J2000}$    & $DEC_{J2000}$  & V     & (B-V)  & $(B-V)_{0}^{1}$  &\multicolumn{2}{c}{ Spectral type} \\
              &              &   (deg)         &   (deg)        & (mag) &  (mag) &  (mag)     & Ph$^2$  & Sp$^3$ \\
\hline         
NGC 869  \#      67 &  LAV 796  &  34.662083       &57.220444   & 14.145$^4$ & 0.445$^4$   & -0.105$^4$     &  B8 & B7$^6$\\
NGC 869  \#     140 &  LAV 1174 &  34.870708       &57.163887   & 11.904$^4$ & 0.407$^4$   & -0.143$^4$     &  B6 & B9$^6$\\ 
NGC 869  \#     42  &  SHM2002 3734   &  34.586708 &57.174110   & 15.733$^4$  & 0.750$^4$  &  0.220$^4$     &  A7 & A7$^6$\\
NGC 869  \#     111 &  BD+56 526$^\star$         &  34.794624       &57.125721   &            &             &                &     &       \\
IC 2602  \#     6   &  V553 Car &  160.249878      &-64.334000  & 15.39$^5$  & 1.570$^5$   &  1.535$^5$     &  M4 &       \\
IC 2602  \#     48  &  V557 Car &  160.72791       &-64.351387  & 10.52$^5$  & 0.600$^5$   &  0.565$^5$     &  G0 & G0$^7$\\ 
\hline
\end{tabular}
\newline
\noindent
$^\dagger$ : The identification number (ID) is taken from Paper I which has been assigned to each X-ray source detected during data reduction procedure\\ 
$^1$ : Intrinsic colors are estimated from observed colors corrected for the extinction (E(B-V)) for their corresponding clusters. For the open clusters NGC 869 and IC 2602, the E(B-V) are taken as
0.55 and 0.035 mag, respectively, from Paper I  \\
$^2$ : Spectral type derived from photometric information in the present study\\
$^3$ : Spectral type derived from spectroscopic studies\\
$^4$ : Data taken from Slesnick, Hillenbrand \& Massey (2002)\\
$^5$ : Data taken from Messina \& Guinan (2003)\\
$^6$ : Spectral type derived from spectroscopic study by Currie et al. (2010)\\
$^7$ : Spectral type derived from spectroscopic study by Glebocki \& Gnacinski (2005)\\
$^\star$ : Most probably a companion/nearby low mass star to this star.\\
\end{table*}

% tab:xlf_test
  %_________________________________________________________________
  \begin{table*}
\caption{Characteristics of flares obtained from the fitting of light curves in energy band 0.3 -- 7.5 keV. FN, $\tau_r$,$\tau_d$, q and A0/q  represent flare number, rising time, decay time, 
count rates at quescient time interval and ratio of peak count rates and count rates at quescient time interval, respectively.   }             % title of Table\label{table:1}      % is used to refer this table in the text
\label{tab:flare_par}      % is used to refer this table in the text
\scriptsize
\begin{tabular}{cccccccc}       % centered columns (4 columns)\hline                 % inserts double horizontal lines
\hline 
Cluster\#ID  &   FN      & Start Time    & Duration &        $\tau_r$     & $\tau_d$        & q                 &  A0/q \\
             &           &    ks         &     ks   &           ks        &   ks            & $\rm{cts~s^{-1}}$ &       \\
\hline         
\hline         
\multicolumn{8}{c}{Intermediate mass stars} \\                     
NGC 869 \#67  &   F1 & 27.0          &  11.0   &  2.33 $\pm$ 0.49 & 1.60 $\pm$ 0.38 & 0.002 $\pm$ 0.001 & 10\\ 
NGC 869 \#140 &   F2 & 0.0           &  10.0   &  1.74 $\pm$ 0.75 & 1.00 $\pm$ 0.31 & 0.003 $\pm$ 0.002 & 7\\ 
\hline         
\multicolumn{8}{c}{Low mass stars}    \\                  
NGC 869 \#42  &   F3&  0.0          &  25.0    &  2.03 $\pm$ 0.72 & 4.62 $\pm$ 0.59 & 0.004 $\pm$ 0.003 & 6.2\\ 
NGC 869 \#111 &   F4& 16.0          &  21.0   &  0.83 $\pm$ 0.14 & 3.00 $\pm$ 0.86 & 0.003 $\pm$ 0.002 & 13.3\\
IC 2602 \#6   &   F5& 24.8          & 13.2     &  0.59 $\pm$ 0.12 & 5.28 $\pm$ 1.70 & 0.014 $\pm$ 0.007 & 7.8\\ 
IC 2602  \#48  &   F6& 18.4          &  9.6     &  1.13 $\pm$ 0.17 & 2.70 $\pm$ 0.25 & 0.349 $\pm$ 0.036 & 2.2\\ 
IC 2602  \#48  &   F7& 28.0          &  8.0     &  2.34 $\pm$ 1.08 & 1.14 $\pm$ 0.87 & 0.349 $\pm$ 0.036 & 1.5\\ 
\hline
\end{tabular}
\end{table*}

% tab:flare_par
  %_________________________________________________________________
  \begin{table*}
\scriptsize
\caption{Spectral parameters of the X-ray flares and quiescent state.} % title of Table
\label{tab:flare_fit_par}      % is used to refer this table in the text
\begin{tabular}{ccccccccc}       % centered columns (4 columns)\hline                 % inserts double horizontal lines
\hline 
Cluster    ID   &FN    & \multicolumn{3}{c}{Flare} &  &\multicolumn{3}{c}{Quiescent State}\\
\cline{2-4} \cline{6-8}
                  &      &$\rm {kT}$              & $\rm{Log(EM))}$     & $\rm{Log(L_X)}$        &             &   $\rm {kT}$              & $\rm{Log(EM))}$     & $\rm{Log(L_X)}$        \\
                  &      &  (keV)           &  ($\rm{cm^{-3}}$)   & $\rm{erg~s^{-1}}$             &             &   (keV)           &  ($\rm{cm^{-3}}$)   & $\rm{erg~s^{-1}}$        \\
\hline 
    NGC869 \#67   & F1     &   $6.50^{+23.90}_{-2.70}$     &   $54.56^{+0.08}_{-0.08}$  & 31.70 &   & $0.54^{+0.43}_{-0.33}$      &    $53.42^{+0.56}_{-0.32}$  & 30.48             \\ 
    NGC869 \#140  & F2     &   $>1.88$                     &   $54.34^{+0.13}_{-0.16}$  & 31.60 &   & $1.27^{+0.43}_{-0.23}$      &    $53.98^{+0.13}_{-0.15}$  & 31.04              \\
    NGC869 \#42   & F3     &   $1.55^{+1.81}_{-0.55}$      &   $54.18^{+0.12}_{-0.21}$  & 31.62 &   & $3.66^{+14.00}_{-1.90}$     &    $54.25^{+0.11}_{-0.12}$  & 31.38              \\
    NGC869 \#111  & F4     &   $7.43^{+8.07}_{-2.65}$      &   $54.47^{+0.05}_{-0.05}$  & 31.66 &   &                             &                             & 30.00$^1$                \\
    IC2602 \#6    & F5     &   $1.95^{+0.68}_{-0.41}$      &   $52.80^{+0.05}_{-0.05}$  & 29.92 &   & $1.31^{+0.31}_{-0.09}$      &    $52.21^{+0.05}_{-0.06}$  & 29.24           \\
    IC2602 \#48   & F6     &   $3.28^{+0.55}_{-0.42}$      &   $53.09^{+0.03}_{-0.03}$  & 30.55 &   & $0.94^{+0.01}_{-0.01}$      &    $53.22^{+0.01}_{-0.01}$  & 30.29     \\
    IC2602 \#48   & F7     &   $2.61^{+0.67}_{-0.46}$      &   $53.04^{+0.05}_{-0.05}$  & 30.53 &   & $0.94^{+0.01}_{-0.01}$      &    $53.22^{+0.01}_{-0.01}$  & 30.29     \\
\hline
\end{tabular}
\newline
\noindent
Notes-- During spectral fitting of data for flare state, the values of $N_H$ have been kept fixed to the quiescent state values in paper I, i.e. 28$\times~10^{20}$ and 1.75$\times~10^{20}$ $cm^{-2}$ for the clusters NGC 869 and IC 2602, respectively. The value of Z is taken as 0.3 solar.\\
$^1$ : X-ray luminosity derived from the count rates in quscient state (see \S\ref{sec:demass}).\\
\end{table*}

% tab:spec_info_flare
  %_________________________________________________________________
  \begin{center}
\begin{table*}
\caption{Loop lengths of flaring loops derived by  using rise and decay method ($L_r$) and Haisch's approach ($L_{Ha}$). Maximum pressure (p), 
plasma density ($\rm{n_e}$), loop volume (V), minimum magnetic field (B) to confine the plasma 
and heating rate per unit volume at the flare peak ($\rm{E_h}$) are also estimated.
}
\label{tab:loop_length} 
\scriptsize     
\begin{tabular}{llllllllll}  
\hline 
Cluster ID &  Flare  &  $\rm{T_{max}^{r}}$&  $\rm{L_{r}}$           &  $\rm{L_{Ha}}$         &    p                           & $\rm{n_e}$        &      V            & B              & $\rm{E_h}$              \\
           &         &                    & ($\rm{10^{10}}$)        & ($\rm{10^{10}}$)       &    ($\rm{10^{3}}$)             & ($\rm{10^{11}}$)  &  ($\rm{10^{30}}$) & ($\rm{10^3}$)  &                         \\
           &         &    MK              &  cm                     &   cm                   &   $\rm{dyne~cm^{-2}}$          &  $\rm{cm^{-3}}$   &  $\rm{cm^3}$      &  G             & $\rm{erg~s^{-1}~cm^{-3}}$\\
\hline  
\multicolumn {10}{c}{Intermediate mass star} \\                                            
NGC 869 \#67  & F1  & $179^{+891}_{-83}$  & $  3.83^{+7.65}_{-1.66} $  &  $ 5.51^{+1.25}_{-1.22} $  &   54.3  &   11.0 &     3.0  &    1.2  &   52.0\\
NGC 869  \#140  & F2  & $>42$               & $  >1.27 $                 &  $ 3.42^{+1.09}_{-1.06} $  &   2.2   &   1.9  &     63.1 &    0.2  &   3.1\\
\hline  
\multicolumn {10}{c}{Low mass stars}           \\                                  
NGC 869   \#42  & F3  & $34^{+49}_{-13}  $  & $  2.93^{+2.67}_{-1.19} $  &  $ 9.84^{+1.71}_{-1.98} $  &   0.5   &   0.5  &  570.2   &    0.1   &   0.3\\
NGC 869   \#111 & F4  & $208^{+281}_{-83}$  & $  3.78^{+3.37}_{-1.55} $  &  $ 8.41^{+2.03}_{-2.08} $  &   87.5  &   15.2 &     1.3  &    1.5  &   91.7\\
IC 2602    \#6  & F5  & $44^{+18}_{-11}  $  & $  2.10^{+1.07}_{-0.77} $  &  $ 4.92^{+1.31}_{-1.35} $  &   1.5   &   1.2  &    4.2   &    0.2   &   1.3\\
IC 2602   \#48  & F6  & $81^{+16}_{-12}  $  & $  3.04^{+0.68}_{-0.56} $  &  $ 3.50^{+0.30}_{-0.30} $  &   6.3   &   2.8  &    1.5   &    0.4   &   5.1\\
IC 2602   \#48  & F7  & $62^{+19}_{-13}  $  & $  3.81^{+2.41}_{-1.96} $  &  $ 1.79^{+1.03}_{-1.20} $  &   3.1   &   1.8  &    3.4   &    0.3   &   2.4\\
\hline
\end{tabular}
\end{table*} 
\end{center}

\end{document}